\documentstyle[aps,preprint]{revtex}

\tightenlines

\begin{document}
\title{Noise effect on the nonadiabatic transition and correction to the tunneling energy gap estimated by the Landau-Zener-St$\rm{\ddot{u}}$ckelberg  formula}

\author{Masamichi Nishino, Keiji Saito, and Seiji Miyashita }

\address{Department of Applied Physics, Graduate School of Engineering,\\ The University of Tokyo, Bunkyoku, Tokyo, Japan}

\maketitle

\begin{abstract}

We study effect of noise on the tunneling process of the magnetization 
in a sweeping field from the viewpoint of the nonadiabatic transition.
The tunneling probability depends on the properties of the noise, e.g. 
the direction, the amplitude, and the relaxation time of the autocorrelation function.
We investigate the tunneling process in the presence of noise by solving the time-dependent Schr$\rm{\ddot{o}}$dinger equation numerically, and analyze its parameter dependence.
We estimate the effective tunneling gap making use of the Landau-Zener-St$\rm{\ddot{u}}$ckelberg (LZS) formula. The dependence of the gap on the sweeping velocity reproduces the corresponding experimental data qualitatively.
For fast sweeping field the estimated gap has no velocity dependence and 
this gap seems to give the true gap. We, however, show that a correction for the estimated energy gap is necessary and we propose an correction formula.

\end{abstract}

\section{Introduction}

The quantum tunneling of the magnetization (QTM) has become one of the new topics in the condensed matter physics.
At low enough temperatures phenomena originating quantum dynamics can be observed in mesoscopic and microscopic systems.
Recently quantum dynamical effects have really been observed in the hysteresis phenomena of the nanoscale molecular magnets such as Mn$_{12}$, Fe$_8$, and V$_{15}$, and the analyses of these phenomena are in progress~\cite{Thomas1,Friedman,Chudnovsky,Hernandez1,Hernandez2,Lionti,Perenboom,Sangregorio,Barbara1,Barbara2,Werns1,Werns2,Werns3,Werns4,Chiorescu,Luis,Fort,Dobrovitski,Gunther}.
These mesoscopic quantum phenomena have been actively investigated not only for academic interests but also for future applications, e.g., in quantum computers.

 We have studied this tunneling dynamics from the viewpoint of the nonadiabatic transition~\cite{Miya1,DeRaedt,Miya2,Saito1,Nishi,Saito2,Miya3}.
The transition probability for the nonadiabatic transition is given by 
the Landau-Zener-St$\rm{\ddot{u}}$ckelberg (LZS) formula in the pure quantum systems, i.e., without noise~\cite{Landau}.
We have shown that the LZS formula is useful for the estimation of the energy gap at the level crossing point.

In the realistic system the effect of the coupling to the environment plays a crucial role in such phenomena.
At low enough temperatures, below thermally activated regime, phonon-mediated relaxation can be neglected and the effects of the ubiquitous hyperfine field within the molecule and the dipole field of the other molecules become significant for the relaxation process. 
They are considered to be the main source of decoherence.
The {\lq}nonexponential relaxation{\rq} phenomena is one of the typical examples.
Time decay of the remanent magnetization measured at low temperatures at the 
early stage has been reported not to obey usual exponential function but shows a square-root function both in the experimental and in the theoretical analyses~\cite{Ohm,Prokof,Cuccoli}.

The LZS tunneling probability is a function of the sweeping velocity and the energy gap at the resonant point.
Because the sweeping velocity and the transition probability can be 
obtained in the experiments, the gap can be estimated from the LZS formula.
In the experiments of Wernsdorfer and co-workers, the tunnel gap has been estimated for molecular magnet Fe$_8$~\cite{Werns1,Werns2,Werns3}. 
The estimated energy gap shows a dependence on the sweeping rate due 
to the noise effects, which does not take place in the pure system.
The estimated gap, however, is almost independent of the sweeping velocity in the range of fast sweeping velocity while it decreases when the velocity 
decreases. 
The estimated gap in the fast sweeping range seems to give a true tunneling energy gap, judging from the fact that the velocity dependence does not exist in the pure quantum system.
Namely the noise effect seems to be negligible in the fast sweeping range. 
We clarify that this is not always true and show that a correction of the estimated gap is necessary. We propose the relation between the true value of the gap and the estimated one. 
   
In this paper we provide a stochastic noise field and investigate the time evolution of the system by solving the Schr$\rm{\ddot{o}}$dinger equation numerically. 
Kayanuma has given the  analytical formulas of the transition probability in the stochastic noise for special cases, and also given numerical results obtained by Wiener-Hermite expansion~\cite{Kaya1}.  The effect of the boson field has been also studied and the transition probabilities have been given analytically for various parameter range~\cite{Kaya2}.   Thus far several studies have been performed on the 
noise effect where the noise is parallel to external field. As far as we know, 
however, the effect of noise applied in other directions has been rarely dealt with and also been limited to the special case. Here we also examine more general cases 
and investigate direction dependence of the noise.

From the averaged time evolution we study how the transition probability, as a function of the sweeping velocity, depends on the direction, the relaxation time, and the amplitude of noise.
We also compare our results with the analytical results for the limiting case~\cite{Kaya1} .
Furthermore, we investigate the relationship between the tunneling gap estimated from fast sweeping rate and the true value, and propose  
a correction form of the estimated energy.

The layout of this paper is as follows. In Sec. \ref{method}, we explain the method used in this paper. In Sec. \ref{damping_dependence}, we study the dependence of the transition probability on the damping factor and the amplitude of noise.  In Sec. \ref{velocity_dependence}, we investigate sweeping velocity dependence of the energy gap estimated from the transition probability and propose the correction of the estimated gap by LZS formula.
A summary and discussions are given in Sec. \ref{summary}


\section{method}
\label{method}

Here we consider a two level system in a sweeping field with noise.
The Hamiltonian is given by
\begin{equation}
{\cal H}(t)=2\Gamma S^x-H(t) S^z+{\cal H^{\rm noise}}(t),
\end{equation}
where $H(t)$ is the sweeping field $H(t)$$=$$(-H_0+ct)$ and ${\cal H^{\rm noise}}(t)$ describes the noise.
We investigate the transition probability under the noise applied in the $x$-, $y$-, or $z$-direction,
\begin{equation}
{\cal H^{\rm noise}}(t)=h_x(t)S^x,  \; \;  h_y(t)S^y  \; \; {\rm or} \; \; h_z(t)S^z.
\end{equation}
We solve the time evolution of the time-dependent
Schr$\rm{\ddot{o}}$dinger equation 
(TDSE)
\begin{equation}
i\hbar\frac{\partial}{\partial t}|\Psi (t) \rangle={{\cal H}(t)}|\Psi (t)\rangle, 
\label{eq:sch}
\end{equation}
where $|\Psi (t)\rangle$ denotes the wave function of the spin system at
time $t$.
Hereafter we set $\hbar=1$.
In this time evolution, we make the exponential decomposition $\exp(-i {\cal H} \Delta t) \simeq \exp(-i {\cal H}_x \Delta t) \exp(-i {\cal H}_y \Delta t) \exp(-i {\cal H}_z \Delta t) $, 
where ${\cal H}_x$ depends only on the $x$-direction operator and similarly in the $y$- or $z$-direction. 
We use the rotating operator $U(a \rightarrow b)$ ($a,b={x,y,z}$) which 
rotates the quantum axis from $a$-axis to $b$-axis and make each exponential 
diagonalized. Using this method~\cite{DeRaedt2}, we can also simulate the case that the noise has components in all directions $(h_x,h_y,h_z)$.
The present system is a two-level system whose adiabatic energy levels are shown  in Fig.~\ref{level_cross} (${\cal H^{\rm noise}}(t)=0$) as a function of the magnetic field $H$.
There the avoided level crossing occurs at $H=0$.

As the initial state, we set $H(t=0)=-H_0<0$, and put the system in the ground state for this field. 
For large $H_0 >0$, the initial state is almost the down state ($M_z=-1/2$ ), 
i.e, $|\Psi (t=0)\rangle=|-\rangle$.
Without noise the probability to remain in the ground state after the crossing is given by the Landau-Zener-St$\rm{\ddot{u}}$ckelberg (LZS)~\cite{Landau}  formula, 
\begin{eqnarray}
p &=& |\langle +| \Psi (t=\infty)\rangle | ^2\\
&=&1-{\rm exp}\left(\frac{-2\pi \Gamma^2}{c}\right),
\label{prob}
\end{eqnarray}
which is a function of the sweeping velocity and the energy gap ($\Delta E=2\Gamma$) at the avoided level crossing point.  

When noise is present, it disturbs the quantum process and hence this formula for the probability changes. We will investigate how the probability depends on the noise. We denote the transition probability under the noise field in the $x$-, $y$- and $z$-direction by $p_x$, $p_y$ and $p_z$, respectively.

It is naturally assumed that noise sources such 
as the hyperfine field and the dipole filed independently cause a change of the internal field with some restoring force on the target spin.
The time evolution of this random noise can be given by a Langevin equation (Ornstein-Ulenbeck process),
\begin{equation}
\dot{h}(t)=-\gamma h(t) + \eta(t), 
\label{Langevin}
\end{equation}
where $\eta(t)$ is a white gaussian noise
\begin{equation}
\langle \eta(t) \rangle=0   \;\;\;{\rm and}\;\;\;
\langle \eta(0)\eta(t) \rangle=A^2\delta(t),
\label{white}
\end{equation}
and $\gamma$ is the damping factor due to the restoring force.

This random noise $h(t)$ has a gaussian distribution and 
an exponential-decaying autocorrelation function. 
\begin{equation}
\langle h(t) \rangle=0 \;\;\;{\rm and}\;\;\;
\langle h(0)h(t) \rangle=\frac{A^2}{2\gamma}
\exp(-\frac{t}{\tau}), 
\label{rand_noise}
\end{equation}
where $\tau=1/\gamma$ is the relaxation time. 
Here $\sqrt {A^2/2\gamma}$ is regarded as the amplitude of the noise and abbreviated by $B \equiv \sqrt {A^2/2\gamma}$.
We prepare $h_x(t)$, $h_y(t)$, and $h_z(t)$ independently.

In realistic situations, the noise has components in all directions.
We also study this case, where we let the noise have the same amplitude as 
in the case of one direction, that is, $\langle h_x^2 + h_y^2 + h_z^2 \rangle = B^2$. We take an average over 1000-4000 samples to estimate the physical properties.

\section{Dependence of the transition probability on the noise}
\label{damping_dependence}

In this section we investigate the dependence of the transition probability 
on the damping factor and also on the amplitude of the noise.
The dependence of the noise direction is also investigated.

Here we focus on the case of slow sweeping velocity because the dependence 
on the direction, the amplitude, and also $\gamma$ appears more
clearly than in the fast sweeping case (see next section).
The field is swept from $H(t=0)$=$-5.0$ to $H(t=t_{\rm final})$=$5.0$ for the system of $\Gamma=0.02$. 
Because $H(t=0) \gg \Gamma$, the initial state is down spin state.
We set the sweeping velocity to be very slow ($c$=0.0005), such that the system almost shows  the adiabatic transition, i.e., $p\simeq1$ when $B=0$.

Figures \ref{t_noise} (a) and (b) show the time dependence of the noise of $B$=0.02 at the values of $\gamma=0.001$ and $\gamma=0.32$, respectively. 
These represent the typical slow and  rapid fluctuations.
Figure \ref{p_gamma} (a) shows dependence of the transition probability on the damping factor $\gamma$ keeping the amplitude constant, $B$=0.02, and Fig. \ref{p_gamma} (b)
shows that for the case of $B$=0.005.
In the former parameter set, the energy gap without noise ($2\Gamma$) and the noise amplitude ($B$) have the same order of magnitude while in the latter parameter set, the noise amplitude $B$ is much smaller than the gap.

As we naturally expect, the noise effect is larger for larger amplitude $B$ of the noise.
Although the absolute values of the transition probabilities are different, they have a similar dependence on the damping factor.
The noise  of $x$- or $y$-direction has stronger influence than the noise in $z$-direction. This is because the transverse field causes the transition to the other level directly.  
For very small value of $\gamma$, the noise can be considered to be static and 
$p_x$, $p_y$, and $p_z$ are approaching to a constant as $\gamma \rightarrow 0$.
The probabilities $p_x$, $p_y$, and $p_z$ decrease rapidly around $\gamma=0.02$.
This value corresponds to the inverse of the transition time, 
\begin{equation}
\frac{1}{t_{\rm tr}}= \frac{1}{\frac{\Gamma}{c}}=0.025.
\end{equation}
The transition time is the time scale to pass the avoided cross region.
This indicates that when the time scale of noise fluctuation becomes shorter 
than that of the transition time, the transition probability changes very sensitively regardless of the noise amplitude. 
The probabilities $p_x$ and $p_y$ take the minimum values at $\gamma=0.1-1.0$ while $p_z$ has the minimum value at $\gamma=0.02-0.05$.
The minimum values of $p_x$ and $p_y$ are close to each other and they are smaller than that of $p_z$.
When $\gamma$ becomes very large, the effect of noise becomes small due to a 
kind of motional narrowing. Let us study these dependences in detail.

\subsection{Slow fluctuation limit $\gamma \rightarrow 0$ }

In the region of small $\gamma$, i.e. in the slow fluctuation case, there is a distinguishable difference between $p_x$ and $p_y$, while $p_y$ and $p_z$ approach to each other. This tendency is found in both figures \ref{p_gamma} (a) and (b).
These dependences can be understood by an analysis of the static 
distribution of noise.

When the noise is applied in the $x$-direction the magnitude of transverse field is $\Gamma+h/2$. Thus, $p_x$ is obtained by averaging over the distribution of $h$:
\begin{equation}
p_x= \left \langle p_{\rm LZS} \left(\Gamma \rightarrow \Gamma+h/2 \right) \right \rangle=
\left \langle1-\exp \left( -\frac{2\pi(\Gamma+h/2)^2}{c}  \right) \right \rangle,
\label{pro_x}
\end{equation}
where $\langle \cdot\cdot\cdot   \rangle$ denotes 
\begin{equation}
\langle \cdot\cdot\cdot \rangle=\frac{1}{\sqrt {2 \pi B^2}} \int_{-\infty}^{\infty} dh \; \cdot\cdot\cdot \;  \exp\left(-\frac{h^2}{2B^2}\right).
\label{average}
\end{equation}
Here the notation ($\Gamma \rightarrow X$) means the replacement of 
$\Gamma$ by $X$ in the LZS transition probability $p_{\rm LZS}$.
In the same way, when the noise is applied in the $y$-direction, the total magnitude of transverse field is $\sqrt{\Gamma^2+(h/2)^2} $, and we have
\begin{equation}
p_y=  \left \langle p_{\rm LZS} \left(\Gamma \rightarrow \sqrt{\Gamma^2+(h/2)^2} \right) \right \rangle=
\left \langle1-\exp \left( -\frac{2\pi(\Gamma^2+(h/2)^2)}{c}  \right) \right \rangle.
\label{pro_y}
\end{equation}
When the noise is applied in $z$-direction,
the field shifts from $H(t)$ to $H(t)+h$ and has no effect on the transition 
probability,
\begin{equation}
p_z=  \left \langle p_{\rm LZS} \right \rangle =\left \langle1-\exp \left( -\frac{2\pi \Gamma^2}{c}  \right) \right \rangle.
\label{pro_z}
\end{equation}

Explicitly evaluating the average with Eq.(\ref{pro_x},\ref{pro_y},\ref{pro_z}), we have 
\begin{equation}
p_x=1-\sqrt{\frac{1}{2 B^2 D} } \exp \left(\frac{1}{D} \left(\frac{\pi}{c}\Gamma \right)^2-\frac{2\pi}{c}\Gamma^2      \right)
\label{px}
\end{equation}
\begin{equation}
p_y=1-\sqrt{\frac{1}{2 B^2 D} } \exp \left(-\frac{2\pi\Gamma^2}{c}  \right), 
\label{py}
\end{equation}
and
\begin{equation}
 p_z=1- \exp \left(-\frac{2\pi\Gamma^2}{c}  \right),
\label{pz}
\end{equation}
where we define
\begin{equation}
D\equiv \frac{\pi}{2c} +\frac{1}{2B^2}.
\end{equation}

When $\Gamma=0$, the transition probability $p_x$ ($=p_y$) has been obtained analytically in the limit of $\gamma \rightarrow 0$ calculating by the infinite series expansion~\cite{Kaya1}. 
This $p_x$ coincides exactly with the special case ($\Gamma=0$) of (Eq. \ref{px}),
\begin{equation}
p_x= \left \langle1-\exp \left( -\frac{2\pi(h/2)^2}{c}  \right) \right \rangle=
1-\left ( 1+\frac{\pi B^2}{c} \right)^{-\frac{1}{2}}.
\end{equation}

Here we find that $p_y > p_z$ and $p_y > p_x$, 
because 
\begin{equation}
2B^2D=2B^2 \left(\frac{\pi}{2c} + \frac{1}{2B^2} \right) >1,
\end{equation}
and 
\begin{equation}
p_y-p_x=\sqrt{\frac{1}{2 B^2 D} } \exp \left(-\frac{2\pi\Gamma^2}{c}  \right) \left (
\exp \left(\frac{1}{D} \left(\frac{\pi}{c}\Gamma \right)^2 \right ) -1 \right ) >0. 
\end{equation}
For the case (a) ($\Gamma=0.02$, $B=0.02$, $c=0.0005$), 
we have $p_x=0.8724$, $p_y=0.9965$, and $p_z=0.9934$, and 
for the case (b) ($\Gamma=0.02$, $B=0.005$, $c=0.0005$), we have 
$p_x=0.9879$, $p_y=0.9939$, and $p_z=0.9934$.
As shown in Figs. \ref{p_gamma} (a) and (b) the values obtained by simulations approach these values.

\subsection{Fast fluctuation limit $\gamma \rightarrow \infty$ }

As $\gamma$ becomes very large, the noise in Eq. (\ref{rand_noise}) becomes a kind of the white noise 
\begin{equation}
\langle h(0)h(t) \rangle=\frac{2B^2}{\gamma} \delta(t) \;\; (\gamma \rightarrow \infty),
\end{equation}
but its amplitude becomes zero
\begin{equation}
\lim_{\gamma \rightarrow \infty} \frac{2B^2}{\gamma}   \int_{0}^{\infty}  \delta(t)  dt =\gamma^{-1} B^2 \rightarrow 0.
\end{equation}
Therefore the effect of noise becomes small and 
the probabilities increase again and reach the probability of LZS regardless of the direction of noise. 
This effect resembles to the motional narrowing which often appears in experiments of spin resonance. Namely, the change by the noise is averaged out.

On the other hand, under the condition that $B^2/\gamma \gg1 $,
$p_z$ in the limit $\gamma \rightarrow \infty$ has been given analytically by Kayanuma~\cite{Kaya1} 
\begin{equation}
p_z=\frac{1-\exp (-4\pi \Gamma^2 /c)}{2}.
\end{equation}
This formula yields $p_z=0.500$ for the parameters $\Gamma=0.02$ and $c=0.0005$. Thus the value of $p_z$ is expected to change from 
the value of LZS to 1/2 when $B=\sqrt{A^2/2\gamma}$ increases.
Let us study how the probability changes as $A$ becomes large for large $\gamma$(=100).
$A$-dependence of $p_z$ is shown in Fig. \ref{A-dependence}, where $p_z$ approach 0.5 as $A$ becomes large as we expect.
The fact that $p_z=0.5$ suggest that when the noise is large, 
large energy pours into the system during the crossing and 
the distribution over the ground state and the excited state equalizes. 
These situations are similar to those with thermal bath at very high temperatures.
We also investigate $A$-dependence of $p_x$ and $p_y$ (Fig. \ref{A-dependence})
and we find $p_x$ and $p_y$ approach 0.5 more rapidly than $p_z$ as $A$ becomes large.

For the special parameter ($\Gamma=0$), $p_x$ in the limit $\gamma \rightarrow \infty$ has been also given analytically by Kayanuma~\cite{Kaya1},
\begin{equation}
p_x(\Gamma=0)=\frac{1-\exp (-\pi B^2 /c)}{2}.
\end{equation}
If we perform a simulation for large values of $\gamma$, 
$p_x$ does not follows this formula, but it approaches 0.
For example we find $p_x(\Gamma=0) \simeq 0$ for ($B=0.02$, $c=0.0005$, and $\gamma=100$), while the above formula gives $p_x(\Gamma=0)=0.459$. 
 The numerical result is due to the abovementioned narrowing effect.
Where does this inconsistency come from?
In analytical treatment, the field is swept from $-\infty$ to $\infty$, while in the simulation the field is swept in a finite region, i.e., from $-H_0$ to $H_0$.
If we set range of sweeping field to be finite in the Kayanuma's derivation, it is found that $p_x(\Gamma=0)=0$~\cite{Kaya3}. Thus the limits of $\gamma \rightarrow \infty$ and $H_0\rightarrow \infty$ are not exchangeable.
A detailed analysis of these limiting cases will be given elsewhere.

\section{ Sweeping velocity dependence of the effective gap}
\label{velocity_dependence}

As we mentioned in the introduction, the sweeping velocity and the transition probability are observed experimentally and the gap is estimated by making use of the LZS formula.
In this section, we investigate the dependence of the estimated energy gap ($\Delta E$) on the sweeping velocity. 

In the absence of noise, the gap at the crossing is given by $\Delta E_0 = 2\Gamma$.
Similarly in the case with noise, the effective gap can be estimated using LZS formula as
\begin{equation}
\Delta E_{\rm eff}=
\sqrt{-\frac{2c}{\pi} \ln (1-p)  }.
\end{equation}

Let us define the following ratio,
\begin{equation}
\Delta \equiv \frac{\Delta E_{\rm eff}}{\Delta E_0}=\frac{1}{2\Gamma} \sqrt{-\frac{2c}{\pi} \ln (1-p)  },
\label{delta}
\end{equation}
 which indicates the strength of the noise effect.
If this ratio is nearly equal to 1, the effect of noise is negligible.
Here we investigate how the noise effect is changed due to the sweeping velocity.

We show the sweeping rate dependence of the effective gap $\Delta E_{\rm eff}$ for various damping factors in Fig. \ref{p_sweep}. Here we set the parameters $B=0.02$. Figures \ref{p_sweep} (a)-(c) show the cases when the noise is applied in $x$-, $y$-, and $z$-direction, respectively.
Figure \ref{p_sweep} (d) shows the case when the noise has components in all directions $(h_x, h_y, h_z)$. Here we let the noise have the same amplitude as in the one-direction case, i.e. $\langle h_x^2 + h_y^2 + h_z^2 \rangle = B^2$. 
The following tendencies appear for the cases (a)-(d).
For large sweeping velocity, the effective gap is found to be constant.
In the range of slow sweeping velocity, the effective gap deviates from this 
constant and decreases roughly linearly with the logarithm of the sweeping velocity.
These results are consistent with the experimental results~\cite{Werns2,Werns3}, where the estimated gap $\Delta E_{\rm eff}$ is constant 
at larger velocity and exhibits a deviation in the low sweeping velocity range.
 We find that this deviation also significantly depends on the damping factor.
As the sweeping velocity gets slow, the system stays near the avoided crossing point for a long time, where the adiabatic motion is largely disturbed, and then effectively the gap becomes small. 

Let us consider the dependence of $\Delta E_{\rm eff}/\Delta E_0$ on $c$ and $\gamma$.
The field crosses the avoided level 
crossing region in a time of the order of $\tau_{\rm tr} \equiv \Gamma/c$. 
When $\gamma$ is much smaller than $\tau_{\rm tr}^{-1} = \gamma_{\rm tr}$, the noise can be regarded to be static. 
We can estimate $\Delta E_{\rm eff}/\Delta E_0$ in the static limit
by using the static noise results for $p_x$, $p_y$, and $p_z$ in Eqs. (\ref{px}), (\ref{py}), and (\ref{pz}).
Substituting them into Eq. (\ref{delta}), we have the velocity dependence of $\Delta E_{\rm eff}^d/\Delta E_0$ ($d$ = $x$, $y$, or $z$).
These dependences are shown in Fig. \ref{Delta_lim} and also shown by the solid lines in Fig. \ref{p_sweep} (a)-(c). We find that these solid lines agree well with the data for large sweeping velocity $c$. 
The shape of the solid line in each direction is different from one another.

When $\gamma$ increases,  the deviation of the effective gap from the solid line becomes large in the slow sweeping 
region and the critical value of the velocity, below which the 
$\Delta E_{\rm eff}^d/\Delta E_0$  ($d$ = $x$, $y$, or $z$) deviates, increases. 
The effect of noise is found to be less important for $z$-direction.
The data of $\gamma=0.32$ and 1.0 in Fig. \ref{p_sweep} (c) are exceptional and  these points represent the behavior in the 
narrowing region in Fig. \ref{p_gamma}, that is, in the region where the transition
probability recovers. Thus the noise is inefficient there and $\Delta_{\rm eff}$ 
behaves differently.

At large $c$, $\tau_{\rm tr}= \gamma_{\rm tr}^{-1} $ is small thus $\gamma_{\rm tr}$ is large. Therefore, even the noise with large $\gamma$s can be regarded to be static and these data still follow the solid line.
If we assume that the noise in the realistic system has a certain 
finite relaxation time, the transition probability corresponds to that of static limit case for very fast sweeping.

We know that $\Delta E_{\rm eff}$ has no sweeping velocity dependence 
for fast sweeping but its value can be different from the true value of the gap.
Let us estimate the correction of the gap.
When the sweeping velocity is large enough, i.e. $2\pi \Gamma^2/c \ll 1$, 
the LZS probability is approximately given by 
\begin{equation}
p_{\rm LZS} \simeq \frac{2\pi  \Gamma^2}{c}=\frac{\pi}{2c}\Delta E_0^2.
\label{LZS_lim}
\end{equation}
Taking the same limit ($\gamma \rightarrow 0$) in the formulas (\ref{px}), (\ref{py}), and (\ref{pz}), 
we have 
\begin{equation} 
p_x= \left \langle p_{\rm LZS} \left(\Gamma \rightarrow \Gamma+\frac{h}{2} \right) \right \rangle=\frac{2\pi \left(\Gamma^2 + \frac{B^2}{4} \right) }{c}
=\frac{\pi}{2c}(\Delta E_{\rm eff}^x)^2,
\label{px_lim}
\end{equation}

\begin{equation}
p_y=  \left \langle p_{\rm LZS} \left(\Gamma \rightarrow \sqrt{\Gamma^2+(h/2)^2} \right) \right \rangle=\frac{2\pi \left(\Gamma^2 + \frac{B^2}{4} \right) }{c}=\frac{\pi}{2c}(\Delta E_{\rm eff}^y)^2,
\end{equation}
and
\begin{equation}
p_z=\left \langle p_{\rm LZS} \right \rangle= \frac{2\pi \Gamma^2 }{c}.
\end{equation}

Thus from $p_z$ we can obtain the true value of the gap $\Delta E_0=2\Gamma$. 
On the other hand, from Eqs. (\ref{LZS_lim}) and (\ref{px_lim}), we have
\begin{equation}
\left(\frac{\Delta E_{\rm eff}^x}{\Delta E_0}\right)^2 =1+\frac{B^2}{4\Gamma^2}.
\end{equation}
Probability $p_y$ has the same correction.
For the present case $B^2/\Gamma^2=1$, we have $\Delta E_{\rm eff}^x/\Delta E_0$=$\Delta E_{\rm eff}^y/\Delta E_0$=1.118, and $\Delta E_{\rm eff}^z/\Delta E_0$=1, which agree with the simulation results (Fig. \ref{p_sweep} (a)-(c)).

For the isotropic static noise, transition probability is given by
\begin{equation}
\bar{p}\equiv \frac{p_x+p_y+p_z}{3}=\frac{2\pi \Gamma^2}{c} + \frac{B^2}{3c}.
\end{equation}
Then we have 
\begin{equation}
\frac{\bar{p}}{p_{\rm LZS}}=
\left(\frac{\Delta \bar{E}_{\rm eff}}{\Delta E_0}\right)^2 =1+\frac{B^2}{6\Gamma^2},  
\end{equation}
that is
\begin{equation}
\Delta \bar{E}_{\rm eff}=\sqrt{1+\frac{B^2}{6\Gamma^2}} \Delta E_0.
\end{equation}
Thus we find that the true gap $\Delta E_0$ is obtained within a correction of $O (\frac{B^2}{\Gamma^2})$.
Even if the noise amplitude is large, the estimated value is proportional to the true value as far as the properties of noise are not changed.
In the present case, because $B^2/\Gamma^2=1$, 
$\Delta \bar{E}_{\rm eff}/\Delta E_0$=1.083, 
which also agree with the simulation results Fig. \ref{p_sweep} (d). 
The effective gaps are always larger than the true value and these give the upper limit of the true gap.

\section{Summary and Discussion}
\label{summary}

We have studied the effect of noise on the nonadiabatic transition in 
the sweeping field.  
The transition probability depends on the amplitude and the relaxation time of the noise.
The probability also depends on the direction of the noise. 
In the slow fluctuation limit of the noise, we can regard the noise to be static, where we obtain the analytical formulas of $p_x$, $p_y$, and $p_z$, and find the relations $p_y>p_z=p_{\rm LZS}$ and $p_y>p_x$. 
When the time scale of the noise fluctuation comes close to that of the transition time, the transition probability drastically changes.
When $\gamma$ increases further, the probabilities approaches that 
of LZS.  There the time scale of the noise fluctuation is much smaller than the transition time. 
We found that there exists an important correlation between these two time scales and the effect of noise.

In the regime of slow sweeping, the estimated gap as a function of the sweeping 
velocity depends on the direction and the damping factor of the noise.
When the sweeping field becomes large, the estimated gap gets constant regardless 
of the damping factor of the noise. It seems that the noise has little effect in this regime. However the effect of noise does not disappear.
These constant values depend on the direction of the noise, 
$\Delta E_{\rm eff}^x=\Delta E_{\rm eff}^y > \Delta E_{\rm eff}^z=\Delta E_{\rm true}$. When the amplitude of noise is not small, the noise effect is not negligible.
We obtained the relationship between the estimated gap and the true one. 

In the Wernsdorfer and co-worker's experiments,  Fe$_8$ 
clusters show a little difference of the estimated gaps between the isotropic modified samples~\cite{Werns2,Werns3}.
These differences may come from the abovementioned noise effect.
In order to clarify the situation, it is necessary to deal with the realistic model ($S=10$) and to estimate the gap for $M_z=\pm 10$, where the effect of noise in the $x$- and $y$-direction would be a little different from the present case.

We believe that our study captures an essence of the noise effect in the relaxation process of spin dynamics and gives important knowledge to following studies.
We will investigate details of a more realistic model in the near future.
We hope that this work provides a basic information for the manipulation 
of microscopic or nanoscale magnetic devices in the future.

\acknowledgements

The authors would like to thank Professor Kayanuma for his useful discussion and to thank Professor Hans De Raedt for his kind critical reading of the manuscript.
The present work was supported by Grant-in-Aid for Scientific
Research from Ministry of Education, Science, Sports and
Culture of Japan.
M. N. is also supported by the Research Fellowships of the Japan 
Society for the Promotion of Science for Young Scientists.

\begin{figure}
\caption{Avoided level crossing.}
\label{level_cross}
\end{figure}

\begin{figure}
\caption{Time dependence of the noise. (a) the slow fluctuation case $(\gamma=0.001)$ and (b) the rapid fluctuation case $(\gamma=0.32)$.}
\label{t_noise}
\end{figure}

\begin{figure}
\caption{ $\gamma$ dependence of the transition probability in the case of 
slow sweeping velocity. 
(a) $B$=0.02  (b) $B$=0.005}
\label{p_gamma}
\end{figure}

\begin{figure}
\caption{$A$-dependence of $p_x$, $p_y$, and $p_z$ for large $\gamma(=100)$.}
\label{A-dependence}
\end{figure}

\begin{figure}
\caption{Field sweeping velocity dependence of the normalized gap.
The noise is applied in each direction for (a) $x$-direction, (b) $y$-direction, and (c) $z$-direction.
The solid line is the slow fluctuation limit of the noise.
The noise has all components for (d).}
\label{p_sweep}
\end{figure}

\begin{figure}
\caption{Field sweeping velocity dependence of the normalized gap for the 
slow fluctuation limit of the noise.}
\label{Delta_lim}
\end{figure}

\end{document}